\documentclass[12pt]{article}
\usepackage[english]{babel}
\usepackage{eucal}
\newcommand{\Li}{\mbox{Li}}
\newcommand{\SN}{\mbox{S}}
\newcommand{\Sf}{\mbox{S}_{1,2}}
\newcommand{\Mvec}{\mbox{\rm\bf M}}
\newcommand{\Nvec}{\mbox{\rm\bf N}}
\newcommand{\tra}{\mbox{Li}_{4}\left(\frac{1}{2}\right)}

\newcommand{\Df}{\stackrel{\tiny\rm Def}{=}}
\newcommand{\beq}{\begin{equation}}
\newcommand{\eeq}{\end{equation}}
\newcommand{\bea}{\begin{eqnarray}}
\newcommand{\eea}{\end{eqnarray}}

\newcommand\ds{\displaystyle}

\newcounter{lin}

\begin{document}
\begin{titlepage}

\begin{flushleft}
DESY 98--141 \hfill {\tt hep-ph/9810241} \\
September 1998                         \\
\end{flushleft}

\vspace{3cm}
\begin{center}
{\LARGE\bf Harmonic Sums and Mellin Transforms}

\vspace{3mm}
{\LARGE\bf up to two--loop  Order}

\vspace{4cm}
{\large Johannes Bl\"umlein and Stefan Kurth}

\vspace{2cm}
{\large\it DESY--Zeuthen, Platanenallee 6, D--15735 Zeuthen, Germany}\\

\vspace{3cm}
\end{center}
\begin{abstract}
\noindent
A systematic study is performed on the finite harmonic sums up to level
four. These sums form the general basis for the Mellin transforms of all 
individual functions $f_i(x)$ of the momentum fraction $x$ emerging in 
the quantities of massless QED and QCD up to two--loop order, as the 
unpolarized and polarized splitting functions, coefficient functions,
and hard scattering cross sections for space and time-like momentum 
transfer. The finite harmonic sums are calculated explicitly in the 
linear representation. Algebraic relations connecting these sums are 
derived to obtain representations based on a reduced set of basic 
functions. The Mellin--transforms of all the corresponding Nielsen 
functions are calculated.
\end{abstract}

\vspace{5mm}
\begin{center}
\end{center}

\end{titlepage}

\section{Introduction}
\label{sec:introd}

\vspace{2mm}
\noindent
For the study of the scaling violations of deep-inelastic scattering
structure functions and other hard scattering processes different
techniques were developed~\cite{TEC,MEL}. Due to mass factorization
the processes can in general be described by a {\sc Mellin}-convolution
of the parton densities and coefficient functions or the hard scattering
cross sections in the CM sub-system, respectively.
The {\sc Mellin}-convolution of two functions $A(x)$ and $B(x)$ is
described by
\begin{equation}
\label{eq0}
\left[A \otimes B\right](x) = \int_0^1 dx_1 \int_0^1 dx_2 \delta(x - x_1
x_2) A(x_1) B(x_2)~.
\end{equation}
In most of the approaches the cross sections are expressed using the
above relation. In the case of the evolution equations of the parton
densities one has to solve coupled integro-differential equations in
$x$-space~\cite{TEC}. On the other hand one may consider the
integral-transformation~\cite{MELI}
\begin{equation}
\label{eq1}
\Mvec[f_i](N) =\int_0^1 dx x^{N-1} f_i(x)~,
\end{equation}
for the functions $f_i(x)~\epsilon~{\cal C}_0^{\infty}]0,1[$,
cf.~\cite{TRIE}, for which $\Mvec[f_i]$ exists.
Here $N$ denotes the integer moment-index.

Since
\begin{equation}
\label{eq2}
\Mvec[A \otimes B](N) = \Mvec[A](N) \cdot \Mvec[B](N),
\end{equation}
the convolutions are mapped into ordinary products leading to
considerable simplifications in most of the problems~\cite{MEL}.
The {\sc Mellin}--transforms $\Mvec[f_i](N)$ appear as the
{\it genuine} quantities in the operator product expansion at integer
$N$. The quantities $\Mvec[A](N)$ may be analytically continued to
complex values of the argument.
Complicated convolution--relations, which emerge in some
applications, can be handled easily by applying the transform 
Eq.~(\ref{eq1}), cf.~\cite{BRN1}. Fast evolution codes operating
at high numerical precision are based on this prescription~\cite{MEL}.
Specific resummations are most easily implemented via the
{\sc Mellin}--representation~\cite{RESU,BV1} allowing also for a thorough
account of the commutation relations between the singlet
matrices~\cite{EKL,BV1}, which is more difficult to be established
in the $x$--space
representation. The procedure can also be applied for analyses of the
hadronic final states~\cite{HV}.

For the subsequent description it appears useful to enlarge the linear
space ${\cal C}_0^{\infty}]0,1[$ to $\widehat{{\cal C}}_0^{\infty}]0,1] =
{\cal C}_0^{\infty}]0,1[ \cup \{\theta(1-x), \delta^{(k)}(1-x),
k~\epsilon~{\rm\bf N}\}$. Here $\theta(1-x)$ and $\delta(1-x)$
denote the {\sc Heaviside}--function and 
$\delta$--distribution~\cite{DIST}, respectively.
The Mellin convolution $\otimes$, Eq.~(\ref{eq0}), induces a 
multiplication in 
$\widehat{{\cal C}}_0^{\infty}]0,1]$~~\footnote{For similar 
considerations concerning other integral transforms, see 
e.g.~\cite{YOSH}.}. This operation is mapped onto ordinary multiplication
in the linear space of the {\sc Mellin}-transforms.

The functions $f_i(x)$ emerging in perturbative calculations in massless
Quantum Field Theories belong to the class discussed by
{\sc Nielsen}~[12--14] and their {\sc Mellin}--convolutions. By explicit
calculation we will show that the {\sc Mellin}--transforms of these
functions can be represented by linear combinations of the finite
harmonic sums
\begin{equation}
\label{eq3}
S_{k_1, \ldots, k_m}(N) =
\sum_{n_1=1}^N \frac{({\rm sign}(k_1))^{n_1}}{n_1^{|k_1|}}
\sum_{n_2=1}^{n_1} \frac{({\rm sign}(k_2))^{n_2}}{n_2^{|k_2|}} \ldots
\sum_{n_m=1}^{n_{m-1}} \frac{({\rm sign}(k_m))^{n_m}}{n_m^{|k_m|}}~,
N~\epsilon~{\bf N},~\forall~l,~k_l~\neq~0.
\end{equation}
Up to the two--loop level harmonic sums up to level
$\lambda = \sum_{j=1}^m |k_j| = 4$ contribute. The finite harmonic sums
are related to the multiple $\zeta$--values\footnote{The notion
derives from the case of single-parameter sums
$\sigma_k = \sum_{n=1}^{\infty} 1/n^{|k|} \equiv \zeta(k)$ for $k > 1$
and $\zeta$ denoting the Riemann $\zeta$-function. Most applications
discuss {\it positive} indices $k_i$ and refer to a somewhat different
ordering in nesting the sums by demanding $n_{i-1} > n_i$ which, however,
is linearly related to the treatment in the present paper. Because
alternating sums do emerge in most of the applications in Quantum
Field Theories we refer to this generalized case subsequently.}
[15--17],
$\sigma_{k_1, \ldots, k_m},~k_1~\neq~1$ in the limit
$N \rightarrow \infty$.

Besides the linear representations various algebraic relations connect the
different harmonic sums.
A still higher symmetry is obeyed by the multiple
$\zeta$--values~\cite{BG1,DB1}.
A comparison of the linear representations with those induced by the
algebraic relations may yield rather non-trivial relations between
{\sc Nielsen}-functions~\cite{JB1}. The systematic classification of the
finite harmonic sums appears to be carried out in a somewhat easier way
than the {\sc Mellin}--transforms of the various {\sc Nielsen}--functions
themselves which is due to the possibility to choose different arguments
and linear combinations of convolutions leading to equivalent
representations in the latter case. Also the algebraic structures are
more easily recognized for the harmonic sums.

It is the aim of the present paper to derive the {\sc Mellin}-transforms
of all individual functions $f_i(x)$ emerging in the description of
the splitting and coefficient functions in massless QED and QCD up to
two--loop order. This also includes the coefficient functions for
time--like processes, such as fragmentation functions, the
{\sc Drell--Yan} process and similar reactions. We investigate their
relations to the finite harmonic sums and derive algebraic relations
among them. They allow for a compact representation of all
{\sc Mellin}--transforms over a fairly reduced set of basic functions.

The paper is organized as follows.  In section~2 we introduce the basic
notations. The structure of anomalous dimensions and coefficient functions
up to two--loop order in terms of {\sc Nielsen}--functions is discussed
Section~3, where we consider the {\sc Wilson}--coefficients of 
$F_L(x,Q^2)$ and the coefficient function $c_2^{\rm NS}(x,Q^2)$ as
examples.
Section~4 contains the explicit linear representations of all harmonic
sums up to level four. Section~5 deals with algebraic relations between 
the finite harmonic sums and section~6 contains the conclusions.
The individual {\sc Mellin}--transforms for the functions $f_i(x)$ 
emerging in the different splitting and coefficient functions are given 
in the appendix.
\section{Harmonic Sums and Mellin Transforms in Field Theory}
\label{sec:harsum}

\vspace{2mm}
\noindent
The coefficient~[21--25]
and splitting
functions~[26--29],
can be expressed as linear
combinations of products of the {\sc Nielsen}--functions [12--14]
\begin{eqnarray}
{\rm S}_{n,p}(x) = \frac{(-1)^{n+p-1}}{(n-1)! p!} \int_0^1 \frac{dz}{z}
\log^{n-1}(z) \log^p(1-zx)~.
\end{eqnarray}
We introduce the notion of the degree of these transcendental functions
as their {\it transcendentality}~$p+n$. The polylogarithms are given by
\begin{eqnarray}
\Li_n(x) = \frac{d \Li_{n+1}(x)}{d \log(x)} = \SN_{n-1,1}(x)~.
\end{eqnarray}
The logarithms $\log(1 \pm x)$ are related to the dilogarithm
by~\cite{GRHO}
\begin{eqnarray}
\frac{d \Li_2(\pm x)}{d \log(x)} = \Li_1(\pm x) = - \log(1 \mp x)
\end{eqnarray}
and
\begin{eqnarray}
\Li_0(x) =  \frac{x}{1-x}~.
\end{eqnarray}
Similarly,
\begin{eqnarray}
\frac{d \SN_{n,p}(x)}{d \log(x)} = \SN_{n-1,p}(x)
\end{eqnarray}
holds. While the logarithms $\log(1 \pm x)$ and $\log(x)$ are of
transcendentality 1, the polylogarithm $\Li_n(x)$ has transcendentality
$n$. The transcendentality of the measures
\begin{eqnarray}
\frac{d x}{1 \pm x}~~~~{\rm and}~~~~\frac{dx}{x}
\end{eqnarray}
is defined to be 1. For general one-dimensional integrals over the
product of a number of {\sc Nielsen}-functions and their derivatives the
transcendentality determines itself
as the sum of the transcendentalities of the
factors of the integrand and the measure.

The simplest harmonic sums $S_{\pm 1}(N)$~\footnote{$S_1(N)$ occurs as 
the
only function being not a polynomial in the 1--loop anomalous dimensions
$\gamma^{(0)}_{qq}$ and $\gamma^{(0)}_{gg}$~\cite{GW}.} are related to
the {\sc Mellin}--transforms
\begin{eqnarray}
\label{eqSa}
S_1(N) &=& \int_0^1 dx \frac{x^N-1}{x-1} \equiv
\Mvec\left[\left(\frac{1}{x-1}\right)_+\right](N) \\
\label{eqSb}
S_{-1}(N) &=& \int_0^1 dx \frac{(-x)^N-1}{x+1} \equiv (-1)^N
\Mvec\left[\left(\frac{1}{x+1}\right)\right](N) - \log(2)~.
\end{eqnarray}
Here the $+$-prescription is defined by
\begin{eqnarray}
\int_0^1 dx g(x) [f(x)]_+ = \int_0^1 dx
\left[g(x) - g(1)\right] f(x).
\end{eqnarray}
The general single harmonic sums $S_{\pm k}(N), k > 0$ are obtained by
the iterated integrals
\begin{eqnarray}
\label{eqSc}
S_k(N) &=&
 \int_0^1 \frac{dx_1}{x_1} \int_0^{x_1} \frac{dx_2}{x_2} \ldots
\int_0^{x_{k-1}}  \frac{x_k^N-1}{x_k-1}
= \frac{(-1)^{k-1}}{(k-1)!} \int_0^1 dx \log^{k-1}(x) \frac{x^N-1}{x-1}
\\
\label{eqSd}
S_{-k}(N) &=&
 \int_0^1 \frac{dx_1}{x_1} \int_0^{x_1} \frac{dx_2}{x_2}
\ldots
\int_0^{x_{k-1}}  \frac{(-x_k)^N-1}{x_k+1}
= \frac{(-1)^{k-1}}{(k-1)!} \int_0^1 dx \log^{k-1}(x)
\frac{(-x)^N-1}{x+1}
\end{eqnarray}
Further alternating or non--alternating summation does naturally lead
to {\sc Mellin}--transforms of {\sc Nielsen}--functions or their
products of higher transcendentality. The  integral representation
of a general harmonic sum $S_{k_l, \ldots, k_1}(N)$ can be obtained
recursively from Eqs.
(\ref{eqSc},\ref{eqSd}) using
\begin{eqnarray}
\label{eqINTR}
\sum_{k=1}^n \frac{x^k}{k^l} &=& \frac{(-1)^{l-1}}{(l-1)!}
\int_0^x dz \log^{l-1}(z) \frac{z^n-1}{z-1}\\
\label{eqINTR4}
\sum_{k=1}^n \frac{(-z)^k}{k^l} &=& \frac{(-1)^{l-1}}{(l-1)!}
\int_0^x dz \log^{l-1}(z) \frac{(-z)^n-1}{z+1}~.
\end{eqnarray}

In the limit $N \rightarrow \infty$ the harmonic sums yield
\begin{eqnarray}
\widetilde{\sigma}_{k_l, \ldots, k_1} = \lim_{N \rightarrow \infty}
S_{k_l, \ldots, k_1}(N)
\label{eqEUS}
\end{eqnarray}
with $\widetilde{\sigma}_{k_l, \ldots, k_1}$ being a polynomial 
of multiple
$\zeta$--values $\sigma_{k_m \ldots k_1}, k_m \neq 1$, and $\sigma_{+1}$,
\begin{eqnarray}
\sigma_{+1} \Df \int_0^1 dx \frac{1}{1-x}~.
\end{eqnarray}
As already the case for the representation of $S_{-1}(N)$,
Eq.~(\ref{eqSb}),  specific numbers as  $\log(2)$ emerge, which
multiply harmonic sums of lower level or {\sc Mellin}--transforms of
functions of lower transcendentality. Although it is not known yet
whether all these constants are transcendental {\it numbers} we apply
the notion of transcendentality introduced above also to them to obtain a
uniform description. Up to transcendentality 4 only a single constant
per level emerges~: $\log(2), \zeta(2), \zeta(3), \Li_4(1/2)$. At higher
levels more constants contribute. For convenience we follow the choice of
Refs.~[31--35].
The constants are at level~5  $\zeta(5)$ and $\Li_5(1/2)$, at level~6 
$\Li_6(1/2)$ and $\sigma_{-5,-1}$, and at level~7 $\Li_7(1/2), 
\sigma_{-5,-1,-1}$ and $\sigma_{5,-1,-1}$, which are believed to be not 
further reducible in terms of $\zeta$-values apart form 
reordering.~\footnote{For  recent reviews on multiple $\zeta$-values
see~\cite{EUREV}.}
\section{Coefficient functions at \boldmath{$O(\alpha_s^2)$}}
\label{sec:coeff}

\vspace{2mm}
\noindent
The {\sc Mellin}--transforms of the splitting functions up to
next--to--leading order~\cite{SPLI2}
 are known for a long time. Most of the
contributing functions can be reduced to polynomials in $S_{\pm k}(N)$
except $\Li_2(x)/(1+x)$ which is related to the sum $S_{-2,1}(N)$,
cf.~section~\ref{sec:dirmellin}~\footnote{See also Ref.~\cite{YN}.}.
Some functions $f_i(x)$ emerge in a combination leading to a simpler
{\sc Mellin}--transform. This is the case for
$\Li_2(-x)+\log(x)\log(1+x)$ which can be expressed by
\begin{equation}
\label{meq2}
\Li_{2}(-x)+\ln x\ln(1+x) = -\frac{1}{2}\widetilde{\Phi}(x)
+\frac{1}{4}\ln^{2}x-\frac{\zeta(2)}{2}~.
\end{equation}
The {\sc Mellin}--transform of the function
\begin{equation}
\widetilde{\Phi}(x) = \int_{x/(1+x)}^{1/(1+x)}\frac{dz}{z}\ln
\left(\frac{1-z}{z}\right)
\end{equation}
itself turns out to be more compact than that of $\Li_2(-x)$ and
$\log(x)\log(1+x)$ individually.

The {\sc Mellin}--transforms of the two--loop coefficient functions
are more complicated than those of the splitting functions. The
simplest among them are the coefficient functions of the longitudinal 
structure function $F_L(x,Q^2)$, which is given by
\begin{eqnarray}
\label{eqfl}
F_L(x,Q^2)\!=\!x\!\left \{
C_{\rm NS}(x,Q^2)\!\otimes\! f_{\rm NS}(x,Q^2)
           + \delta_f\!\left[C_{\rm S}(x,Q^2)\!\otimes\!\Sigma(x,Q^2)
           + C_{\rm g}(x,Q^2)\!\otimes\!G(x,Q^2)\right]\!\right\}
\end{eqnarray}
in the case of pure photon exchange.
The combinations of parton densities are
\begin{eqnarray}
\label{eqpar}
f_{\rm NS}(x,Q^2) &=& \sum_{i=1}^{N_f} e_i^2 \left [q_i(x,Q^2)
+ \overline{q}_i(x,Q^2) \right], \\
\Sigma(x,Q^2) &=& \sum_{i=1}^{N_f}  \left [q_i(x,Q^2)
+ \overline{q}_i(x,Q^2) \right]~.
\end{eqnarray}
$G(x,Q^2)$ denotes the gluon density, $e_i$ the electric charge,
  and $\delta_f = (\sum_{i=1}^{N_f}
e_i^2)/N_f$, with  $N_f$  the number of active flavors.

The coefficient functions $C_i(x,Q^2)$ read
\begin{eqnarray}
\label{eqcoe}
C_{\rm NS}(z,Q^2) &=& a_s   c_{L,q}^{(1)}(z)+
                      a_s^2 c_{L,q}^{(2),NS}(z)  \nonumber\\
C_{\rm  S}(z,Q^2) &=& a_s^2 c_{L,q}^{(2), PS}(z) \\
C_{\rm  g}(z,Q^2) &=& a_s   c_{L,g}^{(1)}(z)+
                      a_s^2 c_{L, g}^{(2)}(z)~,  \nonumber
\end{eqnarray}
where $a_s = \alpha_s(Q^2)/(4\pi)$. The leading order coefficient 
functions are given by~\cite{FLLO}
\begin{eqnarray}
\label{meq3}
 c_{L,q}^{(1)}(z) &=& 4 C_F z  \\
 c_{L,g}^{(1)}(z) &=& 8 N_f z (1-z)~.
\end{eqnarray}
In the $\overline{\rm MS}$ scheme the NLO coefficient functions read
\cite{cl2g2,NZ1}~~\footnote{Previous calculations~\cite{FLNLO2,FLNLO3} 
turned out to be partly incorrect, whereas agreement was shown between
Refs.~\cite{cl2g2,NZ1} and \cite{VL1}. Refs.~\cite{FLNLO3} were later 
corrected in Ref.~\cite{FLNLO4}.}
\begin{eqnarray}
 c_{L,q}^{(2),NS}(z) &=& 4C_{F}(C_{A}-2C_{F})z
 \biggl\{4\frac{6-3z+47z^{2}-9z^{3}}{15z^{2}}\ln z \nonumber \\
& & -4\Li_{2}(-z)[\ln z -2\ln(1+z)]-8\zeta(3)
 -2\ln^{2}z[\ln(1+z)+\ln(1-z)] \nonumber \\
& & +4\ln z\ln^{2}(1+z)-4\ln z\Li_{2}(z) \
 +\frac{2}{5}(5-3z^{2})\ln^{2}z \nonumber \\
& &-4\frac{2+10z^{2}+5z^{3}-3z^{5}}
{5z^{3}}
[\Li_{2}(-z)+\ln z\ln(1+z)] \nonumber \\
& & +4\zeta(2)\left[\ln(1+z)+\ln(1-z)-\frac{5-3z^{2}}{5}\right]
 +8\Sf(-z)+4\Li_{3}(z) \nonumber \\
& &+4\Li_{3}(-z)-\frac{23}{3}\ln(1-z)
 -\frac{144+294z-1729z^{2}+216z^{3}}{90z^{2}}\biggr\} \nonumber \\
& & +8C_{F}^{2}z    \biggl\{\Li_{2}(z)+\ln^{2}z-2\ln z\ln(1-z)
 +\ln^{2}(1-z)-3\zeta(2)\nonumber \\
& & -\frac{3-22z}{3z}\ln z
 +\frac{6-25z}{6z}\ln(1-z)-\frac{78-355z}{36z}\biggr\}
\nonumber \\
& & -\frac{8}{3}C_{F}N_{f}z    \left\{2\ln z -\ln(1-z) -\frac{6-25z}
{6z}\right\},
\\
c_{L,q}^{(2),PS}(z) &=& \frac{16}{9z}C_{F}N_{f}\Bigl\{3(1-2z-2z^{2})
(1-z)\ln(1-z)
 +9z^{2}[\Li_{2}(z)+\ln^{2}z-\zeta(2)]\nonumber \\
& &~~~~ +9z(1-z-2z^{2})\ln z
 -9z^{2}(1-z)-(1-z)^{3}\Bigr\},
\\
\label{meq4}
c_{L,g}^{(2)}(z) &=& C_{F}N_{f}\biggl\{16z[\Li_{2}(1-z)
+\ln z\ln(1-z)] \nonumber \\
& & +\left(-\frac{32}{3}z+\frac{64}{5}z^{3}+\frac{32}{15z^{2}}\right)
[\Li_{2}(-z)+\ln z\ln(1+z)]
 +(8+24z-32z^{2})\ln(1-z) \nonumber \\
& & -\left(\frac{32}{3}z+\frac{32}{5}z^{3}\right)
\ln^{2}z +\frac{1}{15}\left(-104-624z+288z^{2}-\frac{32}{z}
\right)\ln z \nonumber \\
& & +\left(-\frac{32}{3}z+\frac{64}{5}z^{3}\right)\zeta(2)
-\frac{128}{15}-\frac{304}{5}z+\frac{336}{5}z^{2}+\frac{32}{15z}\biggr\}
\nonumber \\
& & +C_{A}N_{f}\biggl\{-64\Li_{2}(1-z)+(32z+32z^{2})[\Li_{2}(-z)
+\ln z\ln(1+z)] \nonumber\\
& &+(16z-16z^{2})\ln^{2}(1-z)
+(-96z+32z^{2})\ln z\ln(1-z)
\nonumber\\ & &
+\left(-16-144z+\frac{464}{3}z^{2}
+\frac{16}{3z}\right)\ln(1-z)
+48z\ln^{2}z    +(16+128z-208z^{2})\ln z
\nonumber\\  & &
+32z^{2}\zeta(2)+\frac{16}{3}
+\frac{272}{3}z-\frac{848}{9}z^{2}-\frac{16}{9z}\biggr\}~,
\end{eqnarray}
with $C_A = N_c =3, C_F = (N_c^2 - 1)/(2 N_c) = 4/3$.
The corresponding expressions in the DIS scheme are given
in~\cite{cl2g2}. The class of basic functions is the same for both
schemes.

Also in this case some of the functions $f_i(x)$ can be combined
to somewhat simpler {\sc Mellin}--transforms. This is e.g. the case for
the combination
\begin{equation}
8\Li_{2}(-z)\ln(1+z)+4\ln z\ln^{2}(1+z)+8\Sf(-z)
\end{equation}
in $c_{L,q}^{(2),NS}(z)$. These terms can be transformed using the
relation
\begin{eqnarray}
\lefteqn{\int_{0}^{1}dz\,z^{N-1}\bigl[2\Li_{2}(-z)\ln(1+z)
+\ln z\ln^{2}(1+z)+2\Sf(-z)\bigr]} \nonumber \\
& & =\frac{1}{N}\int_{0}^{1}dz\,z^{N}\biggl[\frac{\tilde{\Phi}(z)}{1+z}
-\frac{1}{2}\frac{\ln^{2}z}{1+z}+\zeta(2)\frac{1}{1+z}\biggr]
-\frac{\zeta(2)}{N}\ln 2+\frac{\zeta(3)}{4N}.
\label{newrel}
\end{eqnarray}

In the coefficient function $c_2^{(2){\rm NS}}(x)$~\cite{NZ1}
and similar quantities~[23--25]
even more complicated functions $f_i(x)$ such as
\begin{equation}
\frac{\log(x)}{1+x} \log(1+x) \log(1-x), \nonumber
\end{equation}
or
\begin{equation}
\frac{1}{1+x}   \left[ \Li_3\left(\frac{1-x}{1+x}\right)
 - \Li_3\left(-\frac{1-x}{1+x}\right) \right]
\label{futyp}
\end{equation}
emerge up to two--loop level. The full set of about 80 functions $f_i(x)$
and their {\sc Mellin}--transforms is given in the appendix in tabular
form in terms of the linear representations derived in
section~\ref{sec:dirmellin}. 


\renewcommand{\arraystretch}{1.3}
\begin{center}
\begin{tabular}{||r||r|r|r||}
\hline \hline
\multicolumn{1}{||c}{$N$}&
\multicolumn{1}{||c}{$B_{2,F}(N)$}&
\multicolumn{1}{|c}{$B_{2,A}(N)$}&
\multicolumn{1}{|c||}{$B_{2,N_F}(N)$}\\
\hline \hline
 2 & 17.9078876005 &  $-$3.53561968276 & $-$3.99999999998\\
 4 & 21.4337645358 &     31.8422327506 & $-$12.7409351851\\
 6 & 49.6093136699 &     58.4783326748 & $-$21.0097878018\\
 8 & 76.6379137872 &     92.3938671695 & $-$28.4436170993\\
10 & 101.906494546 &     130.266385188 & $-$35.1459888118\\
12 & 125.462647526 &     170.288348629 & $-$41.2458493431\\
14 & 147.482822156 &     211.433606893 & $-$46.8491181970\\
16 & 168.152110570 &     253.091881557 & $-$52.0380522803\\
18 & 187.635850771 &     294.886826041 & $-$56.8764477704\\
20 & 206.075234490 &     336.580388240 & $-$61.4143760545\\
\hline \hline
\end{tabular}
\renewcommand{\arraystretch}{1}

\vspace{3mm} {\sf Table~1:~Numerical values of the coefficients
$B_{2,j}(N)$ in Eq.~(\ref{c2nsN}).}
\end{center}

\vspace{3mm}
\noindent
All {\sc Mellin}--transforms can be expressed
in terms of linear combinations of up to threefold sums. The fourfold
sums, which emerge in intermediary steps, are found to be reducible to
more elementary sums by algebaric relations, 
cf.~section~\ref{sec:algebra}.
From this representation the {\sc Mellin}--transforms of the different
splitting functions, coefficient functions and hard scattering cross
sections for space and time--like unpolarized and polarized processes in
massless QED and QCD up to two--loop order can be assembled directly.
The
expressions apply for integer values of the {\sc Mellin}--index $N$. The
analytic continuations to complex values of $N$ are derived in
Ref.~\cite{JB3}. For a computer--code containing these quantities see
Ref.~[2c].

The explicit expressions for the {\sc Mellin}--transforms of the
individual two--loop quantities are rather lengthly. Rather than
giving the explicit expressions for the different coefficient functions
we are going to provide the {\sc Mellin}--transforms of the contributing 
functions $f_i(x)$. This allows to calculate also the 
{\sc Mellin}--transform for an even wider class of processes.

An example which covers the more complicated {\sc Mellin}--transforms for 
the functions $f_i(x)$ is the coefficient function 
$c_2^{(2){\rm NS}}(x)$, which was also studied in Ref.~\cite{VERM2} 
recently.

We performed a numerical comparison of our representation with that given
in \cite{VERM2} for the first 20 moments and find agreement, see table~1.
\begin{eqnarray}
c_2^{(2),\rm NS}(N) = C_F^2 B_{2,F}(N) + C_F C_A B_{2,A}(N)
+ C_F N_F B_{2,N_F}(N)~.
\label{c2nsN}
\end{eqnarray}
The numbers were also compared with the first 10 moments given in 
Ref.~\cite{NZ1} and the analytic results in Refs.~\cite{VL1}.
\section{Linear Representations of Harmonic Sums by Mellin Transforms}
\label{sec:dirmellin}

\vspace{2mm}
\noindent
In the following we give explicit representations of all finite
harmonic sums up to level four in terms of a linear combination
between the {\sc Mellin} transform of a {\sc Nielsen}--function of the
same level and harmonic sums of lower level. Contrary to the
algebraic relations being discussed in section~\ref{sec:algebra} we call
these representations linear. They are obtained by consecutive
partial integration starting from the integral representation
Eqs.~(\ref{eqSc}--\ref{eqINTR4}).
The {\sc Mellin} transforms of the individual
functions $f_i(x)$ of the various two--loop quantities given in the 
appendix will be expressed in terms of these harmonic sums. Up to 
two--loop order not all the sums are emerging. They will, however,
contribute to expressions at higher loop level, similarly as in the case 
of lower level sums in the cases discussed below. We add as well the 
analytic continuations for the most simple cases, in which they can be 
expressed by the functions $\psi^{(k)}(z)$ and $\beta^{(k)}(z)$, where
\begin{eqnarray}
\psi(x)  &=& \frac{1}{\Gamma(x)} \frac{d}{dx} \Gamma(x)\\
\beta(x) &=& \frac{1}{2} \left[
\psi\left(\frac{x+1}{2}\right)
          -  \psi\left(\frac{x}{2}\right)\right]~.
\end{eqnarray}
As a shorthand notation we denote by $F_1(x)$ the following
combination of {\sc Nielsen}--functions
\begin{eqnarray}
\label{eqF1}
F_1(x) &=& \Sf\left(\frac{1-x}{2}\right) + \Sf(1-x) - \Sf\left(
\frac{1-x}{1+x}\right) + \Sf\left(\frac{1}{1+x}\right) - \log(2)
\Li_2\left(\frac{1-x}{2}\right) \nonumber\\  & &
+ \frac{1}{2} \log^2(2) \log\left(
\frac{1+x}{2}\right) - \log(2) \Li_2\left(\frac{1-x}{1+x}\right)~,
\end{eqnarray}
which contributes to a series of harmonic sums given below.

The linear representation of the harmonic sums up to level four 
are given by
%
%
\subsection{First Order Sums}
%
\begin{eqnarray}
\label{eqME1}
S_{-1}(N) &=& (-1)^{N}\Mvec\left[\frac{x^{N}}{1+x}\right](N)
 - \log(2)
              \;\;\;\; \nonumber\\
&=& (-1)^{N}\beta(N+1) - \log(2)  \\
%
%
& & \nonumber\\
S_{1}(N) &=& \Mvec\left[\left(\frac{1}{x-1}\right)_{+}\right](N)
            \;\;\;\; \nonumber\\
&=& \psi(N+1) + \gamma_{E}
\end{eqnarray}
%
\subsection{Second Order Sums}
%
\begin{eqnarray}
S_{-2}(N) &=& (-1)^{N+1}\Mvec\left[\frac{\log x}{1+x}\right](N)
                  - \frac{1}{2}\zeta(2)
           \;\;\;\; \nonumber\\
&=&  (-1)^{N+1}\beta'(N+1) - \frac{1}{2}\zeta(2) \\
%
%
& & \nonumber\\
S_{2}(N) &=& -\Mvec\left[\frac{\log x}{x-1}\right](N)
 + \zeta(2)
          \;\;\;\; \nonumber\\
&=&  -\psi'(N+1) + \zeta(2) \\
%
%
& & \nonumber\\
S_{-1,-1}(N) &=& -\Mvec\left[\left(\frac{\log(1+x)}{x-1}\right)_+\right]
(N) + \log(2) \left[S_1(N)-S_{-1}(N)\right]
\nonumber\\
&=& \frac{1}{2} \left\{\left[\beta(N+1)-(-1)^N \log(2)\right]^2
+ \zeta(2) - \psi'(N+1)\right\}
 \\
%
%
& & \nonumber\\
S_{-1,1}(N) &=& (-1)^{N+1} \Mvec\left[\frac{\log(1-x)}{1+x}\right](N)
- \frac{1}{2} \left[\zeta(2) - \log^2(2)\right] \nonumber\\ &=&
(-1)^{N}\Mvec\left[\frac{\log(1+x)}{1+x}\right](N)
                   + S_{1}(N)S_{-1}(N) + S_{-2}(N) \nonumber\\
 & &              + \left[S_{1}(N)
                   - S_{-1}(N)\right]\log(2)
- \frac{1}{2}\log^{2}(2) \\
%
%
& & \nonumber\\
S_{1,-1}(N) &=& (-1)^{N+1}\Mvec\left[\frac{\log(1+x)}{1+x}\right](N)
                  - \left[S_{1}(N) - S_{-1}(N)\right] \log(2)
                  + \frac{1}{2}\log^{2}(2)
\\
%
%
& & \nonumber\\
S_{1,1}(N) &=& -\Mvec\left[\left(\frac{\log(1-x)}{x-1}\right)_+\right]
(N)
\nonumber\\
&=& \frac{1}{2} \left\{\left[\psi(N+1)+\gamma_E \right]^2
+ \zeta(2) - \psi'(N+1)\right\}
\end{eqnarray}
Here $\gamma_E$ denotes the {\sc Mascheroni}--number.
%
\subsection{Third Order Sums}
%
\begin{eqnarray}
S_{-3}(N) &=& (-1)^{N} \frac{1}{2} \Mvec\left[\frac{\log^{2}x}{1+x}
\right](N)
           - \frac{3}{4}\zeta(3)
      \;\;\;\; \nonumber\\
&=& (-1)^{N}\frac{1}{2}
\beta''(N+1) - \frac{3}{4}\zeta(3)\\
%
%
& & \nonumber\\
S_{3}(N) &=& \frac{1}{2}\Mvec\left[\frac{\log^{2} x}{x-1}\right](N)
                 + \zeta(3)
\nonumber\\
&=& \frac{1}{2} \psi''(N+1) + \zeta(3)
\\
%
%
& & \nonumber\\
S_{-2,-1}(N) &=& - \Mvec\left[\left(\frac{\Li_2(-x)}{x-1}\right)_+
\right](N)
 + \log(2) \left[S_2(N)-S_{-2}(N)\right] - \frac{\zeta(2)}{2}
S_1(N)
\\
%
%
& & \nonumber\\
S_{-2,1}(N) &=& (-1)^{N+1}\Mvec\left[\frac{\Li_{2}(x)}{1+x}\right](N)
                  + \zeta(2)S_{-1}(N) - \frac{5}{8}\zeta(3)
                  + \zeta(2)\log(2) \\
%
%
& & \nonumber\\
S_{2,-1}(N) &=& (-1)^{N+1} \Mvec\left[\frac{\Li_2(-x)}{1+x}\right](N)
- \log(2) \left[S_2(N) - S_{-2}(N)\right]
- \frac{1}{2} \zeta(2) S_{-1}(N) \nonumber\\ & &
 + \frac{1}{4} \zeta(3) - \frac{1}{2}
\zeta(2) \log(2)
\\
%
%
& & \nonumber\\
S_{2,1}(N) &=&
\Mvec\left[\left(\frac{\Li_2(x)}{x-1}\right)_{+}\right](N)
                  + \zeta(2)S_{1}(N)  \\
%
%
& & \nonumber\\
S_{-1,-2}(N) &=& \Mvec\left\{\left[\frac{\log(1+x) \log(x) + \Li_2(-x)}
{x-1}\right]_+\right\}(N)+\frac{1}{2}\zeta(2)
\left[S_1(N)-S_{-1}(N)\right]
\\
%
%
& & \nonumber\\
S_{-1,2}(N) &=&
             (-1)^{N+1} \Mvec\left[\frac{\Li_{2}(1-x)}{1+x}\right](N)
              + \zeta(2)S_{-1}(N) - \zeta(3)
              + \frac{3}{2}\zeta(2)\log(2) \\
%
%
& & \nonumber\\
S_{1,-2}(N) &=& (-1)^N \Mvec\left[\frac{\log(x) \log(1+x)+\Li_2(-x)}
{1+x}\right](N) - \frac{1}{2} \zeta(2) \left[S_1(N) - S_{-1}(N)\right]
\nonumber\\ & & - \frac{1}{8} \zeta(3) + \frac{1}{2} \zeta(2) \log(2)
%
\\
%
%
& & \nonumber\\
S_{1,2}(N) &=& -\Mvec\left[\frac{\Li_2(1-x)}{x-1}\right](N) + \zeta(2)
S_1(N) - \zeta(3)
\\
%
%
& & \nonumber\\
S_{-1,-1,-1}(N) &=& (-1)^{N+1} \Mvec\left[\frac{\Li_2[(1-x)/2] - \log(2)
\log(1-x)}{1+x}\right](N) \nonumber\\ & &
+ \log(2) \left[S_{-1,1}(N)-S_{-1,-1}(N)
\right] \nonumber\\  & &
+ \frac{1}{2} \left[\zeta(2) - \log^2(2) \right] S_{-1}(N)
- \frac{1}{4} \zeta(3) + \zeta(2) \log(2) - \frac{2}{3} \log^3(2)
\nonumber\\ &=& \frac{1}{6} \left[(-1)^N \beta(N+1) - \log(2)\right]^3
\nonumber\\ & &
- \frac{1}{2}\left[(-1)^N\beta(N+1)-\log(2)\right]\left[\psi'(N+1)
- \zeta(2)\right]
\nonumber\\ & &
+ \frac{1}{3} \left [(-1)^N \frac{1}{2}
\beta''(N+1) -
\frac{3}{4} \zeta(3) \right]
\\
%
%
& & \nonumber\\
S_{-1,-1,1}(N) &=& \Mvec\left\{\left[\frac{1}{x-1}\left(\log(1-x)
\log\left(\frac{1+x}{2}\right) + \Li_2\left(\frac{1-x}{2}\right)\right)
\right]_+
\right\}(N)  \nonumber\\ & &
- \frac{1}{2} \left[\zeta(2) - \log^2(2)\right] S_{-1}(N)
\\
%
%
& & \nonumber\\
S_{-1,1,-1}(N) &=& \frac{1}{2}\Mvec\left\{\left[\frac{\log^2(1+x)}{x-1}
\right]_+\right\}(N) + \log(2) \left[S_{-1,-1}(N)-S_{-1,1}(N)\right]
\nonumber\\ & &
- \frac{1}{2} \log^2(2)\left[S_1(N)-S_{-1}(N)\right]
\\
%
%
& & \nonumber\\
S_{-1,1,1}(N) &=& \frac{1}{2} (-1)^N
\Mvec\left[\frac{\log^2(1-x)}{1+x}\right](N) - \frac{7}{8} \zeta(3)
+ \frac{1}{2} \zeta(2) \log(2) - \frac{1}{6} \log^3(2) \\ 
%
%
& & \nonumber\\
S_{1,-1,-1}(N) &=& \Mvec\left\{
\left[\frac{1}{x-1}\left(\log\left(\frac{1-x}{2}\right)\log(1+x) +
\Li_2\left(\frac{1+x}{2}\right)\right)
\right]_+\right\}(N)\nonumber\\ & &
+ \log(2) \left[S_{1,1}(N)-S_{1,-1}(N)\right] - \frac{1}{2} \left[
\zeta(2) - \log^2(2) \right] S_1(N)    \\
%
%
& & \nonumber\\
S_{1,-1,1}(N) &=&
(-1)^N \Mvec\left\{\frac{1}{1+x} \left[\log\left(\frac{1+x}{2}\right)
\log(1-x) + \Li_2\left(\frac{1-x}{2}\right)\right]\right\}(N)
\nonumber\\ & &
- \frac{1}{2}\left[\zeta(2)-\log^2(2)\right] S_1(N)
 + \frac{1}{8} \zeta(3)
- \frac{1}{2} \zeta(2) \log(2) +\frac{1}{3} \log^3(2)
\\
%
%
& & \nonumber\\
S_{1,1,-1}(N) &=& (-1)^{N} \frac{1}{2}
\Mvec\left[\frac{\log^2(1+x)}{1+x}\right](N) + \log(2) \left[S_{1,-1}(N)-
S_{1,1}(N)\right] \nonumber\\ & &
+ \frac{1}{2} \log^2(2) \left[S_1(N) - S_{-1}(N)\right]
- \frac{1}{6} \log^3(2)
\\
%
%
& & \nonumber\\
S_{1,1,1}(N) &=& \frac{1}{2} \Mvec\left[\left(\frac{\log^2(1-x)}{x-1}
\right)_+\right](N)
\nonumber\\ &=&
\frac{1}{6} \left[\psi(N+1)+\gamma_E\right]^3
- \frac{1}{2}\left[\psi(N+1)+\gamma_E\right]\left[\psi'(N+1)
- \zeta(2)\right]
\nonumber\\ &=&
+ \frac{1}{3} \left [\frac{1}{2} \psi''(N+1) +
\zeta(3) \right]
\end{eqnarray}
%
\subsection{Fourth Order Sums}
%

%
%

The respective combinations of the linear representations of individual
harmonic sums may be compared with the algebraic representations of
these combinations as given in section~\ref{sec:algebra}. In this way
interesting convolution relations between {\sc Nielsen}-functions
may be obtained, which we do not work out further here.

For the above relations extensive checks have been performed by
thorough numerical precision test and using computer algebra programs.
As a further check we compared the limit
\begin{eqnarray}
\lim_{N \rightarrow \infty} S_{k_1, \ldots, k_l}(N) =
\sigma_{k_1, \ldots, k_l}
\end{eqnarray}
with the
respective  multiple $\zeta$-values in the cases when $k_1 \neq 1$.
Analytic results for $\sigma_{k_1, \ldots, k_l}$ are obtained
straightforwardly from the integral
{\sc Euler--Poincar\'{e}}~\cite{EUPO} integral representation
Eqs.~(\ref{eqSc}--\ref{eqINTR4}) for $N \rightarrow \infty$.
Nearly all level--four multiple $\zeta$--values were given by
{\sc Gastmans} and {\sc Troost}~\cite{GATR}. The missing ones are
\begin{eqnarray}
\sigma_{-1,-3} &=& 2 \Li_4\left(\frac{1}{2}\right) - \frac{1}{2} \zeta(4)
+ \frac{3}{4}\zeta(3)\log(2) - \frac{1}{2} \zeta(2) \log^2(2)
+ \frac{1}{12} \log^4(2) \\
\sigma_{-1,+3} &=&  - \frac{19}{16} \zeta(4) + \frac{3}{4} \zeta(3)
\log(2)~.
\end{eqnarray}
In the integrations extensive use has been made of the integrals for 
{\sc Nielsen}--functions~[31,46--48].
\section{Algebraic Relations between Harmonic Sums}
\label{sec:algebra}

\vspace{2mm}
\noindent
The finite harmonic sums are connected by various algebraic relations.
We will only consider those among the harmonic sums 
themselves~\footnote{For the study of relations between infinite
harmonic sums the introduction of non-harmonic rest terms which vanish 
as $N \rightarrow \infty$ may sometimes be useful~\cite{BG1}.}.
One class of relations is obtained by considering the full set of 
permutations of an index-set and  interchanging the order of summation. 
A second class follows from partial permutations.
\subsection{Complete Index Permutation}
\label{sec:algebra1}
In the case of two indices the relation was given firstly by 
{\sc Euler}~\cite{EULER} for $N \rightarrow \infty$, see 
also~[49--51].
Below we give the generalizations up
to four (alternating) indices and consider also relations for 
structurally more simple harmonic sums beyond transcendentality four. Due
to the index-symmetry the combinations listed below are polynomials of 
the
{\it simple} harmonic sums only, i.e. the 
{\sc Mellin}-transform to $x$--space
of these combinations is a linear combination out of polynomials of
{\sc Mellin}-convolutions of the {\it basic functions}
\begin{eqnarray}
\frac{(-1)^n}{\Gamma(n)} \left(\frac{\log^{n-1}(x)}{x-1}\right)_+
&=& \Mvec^{-1}\left[S_n(N)\right](x) \\
\frac{(-1)^N}{x+1}
- \log(2) \delta(1-x)
&=& \Mvec^{-1}\left[S_{-1}(N)\right](x),  \\
\frac{(-1)^{N+n-1}}{\Gamma(n)} \left(\frac{\log^{n-1}(x)}{x+1}\right)
- \left(1 - \frac{1}{2^{n-1}}\right) \zeta(n) \delta(1-x)
&=& \Mvec^{-1}\left[S_{-n}(N)\right](x), n~\epsilon~\Nvec, n > 1
\nonumber\\
\end{eqnarray}
only. As will be shown below, the permutation--symmetric combinations
obey also nested representations by lower level harmonic sums, which are
not single. By comparing the algebraic relations given below with the 
linear representations of the previous section one may find convolution 
relations between {\sc Nielsen}--functions.

The general permutation--relation for sums with two indices reads
\begin{eqnarray}
S_{m,n} + S_{n,m} &=& S_m S_n + 
S_{{\rm sign}\{m\} {\rm sign}\{n\}
[|m| + |n|]}\nonumber\\
&\equiv& S_m S_n + S_{m \wedge n}~.
\nonumber\\
\label{eqSR2}
\end{eqnarray}
A geometric argument for its derivation is the following. The sums
$S_{m,n}$ and $S_{m,n}$ correspond to congruent rectangular triangles
which may be supplemented to form the rectangle $S_m S_n$. Since both
triangles overlap on the diagonal of the rectangle
$S_{m \wedge n}$ has to be added.

In this way one obtains up to the level of transcendentality four
14 relations of the type~(\ref{eqSR2}), which read
\begin{eqnarray}
S_{1,1}  &=& \frac{1}{2} \left[S_1^2 + S_{2}\right]
\\ [2mm]
S_{1,-1} + S_{-1,1} &=& S_1 S_{-1} + S_{-2}
\\ [2mm]
S_{-1,-1}  &=& \frac{1}{2} \left[S_{-1}^2 + S_{2}\right]
\\ [2mm]
\Sf + S_{2,1} &=& S_1 S_2 + S_3
\\ [2mm]
S_{-1,2} + S_{2,-1} &=& S_{-1} S_2 + S_{-3}
\\ [2mm]
S_{1,-2} + S_{-2,1} &=& S_{1} S_{-2} + S_{-3}
\\ [2mm]
S_{-1,-2} + S_{-2,-1} &=& S_{-1} S_{-2} + S_{3}
\\ [2mm]
S_{2,2}  &=& \frac{1}{2} \left[S_{2}^2 + S_{4}\right]
\\ [2mm]
S_{2,-2} + S_{-2,2} &=& S_{2} S_{-2} + S_{-4}
\\ [2mm]
S_{-2,-2}  &=& \frac{1}{2} \left[S_{-2}^2 + S_{4}\right]
\\ [2mm]
S_{1,3} + S_{3,1} &=& S_1 S_3 + S_4
\\ [2mm]
S_{-1,3} + S_{3,-1} &=& S_{-1} S_3 + S_{-4}
\\ [2mm]
S_{1,-3} + S_{-3,1} &=& S_1 S_{-3} + S_{-4}
\\ [2mm]
S_{-1,-3} + S_{-3,-1} &=& S_{-1} S_{-3} + S_{4}
\end{eqnarray}
Similarly, one obtains for three indices
\begin{eqnarray}
\sum_{{\rm perm}\{l,m,n\}} S_{l,m,n} &=& S_l S_m S_n 
+ \sum_{{\rm inv~perm}\{l,m,n\}} S_{l}
S_{m \wedge n}
+ 2~S_{l \wedge m \wedge n}~,
\label{perm3}
\end{eqnarray}
where {\sf `inv perm'} denotes the invariant permutations and {\sf
`perm'} all permutations.
In a geometric interpretation one may identify the sums $S_{l,m,n}$ with
the six pyramids of equal volume\footnote{Due to the 
solution~\cite{HSOL}  of the 3rd {\sc Hilbert}--problem \cite{HILBERT} 
these pyramids are not congruent in general.}
, which form the square stone
$S_l S_m S_n$. The terms 
$\sum_{{\rm inv~perm}\{l,m,n\}} S_{l} S_{m \wedge n}$ account for 
additional
overlapping surfaces in the interior of the square stone and
$2 S_{l \wedge m \wedge n}$ for overlapping
internal edges. The corresponding relation for infinite sums was
derived in Ref.~\cite{RS1}~\footnote{For related work see 
also~\cite{HOF}.}.

Up to transcendentality four 10 relations are obtained,
\begin{eqnarray}
S_{-1,-1,-1} &=& \frac{1}{6} S_{-1}^3 + \frac{1}{2} S_{-1} S_2 
+ \frac{1}{3} S_{-3} \nonumber\\ &=&
\frac{1}{3} \left[S_{-1,-1} S_{-1} + S_{-1} S_2 + S_{-3}\right]
\\ [2mm]
S_{-1,-1,1} + S_{-1,1,-1} + S_{1,-1,-1}
 &=& \frac{1}{2} \left[S_{-1}^2 + S_2\right] S_1 + S_{-1} S_{-2}
+ S_3 \nonumber\\ &=&
S_{-1,-1} S_1 +S_{-1} S_{-2} + S_3
\\ [2mm]
S_{-1,1,1} + S_{1,-1,1} + S_{1,1,-1}
&=& \frac{1}{2} \left[S_1^2 + S_2 \right] S_{-1} + S_1 S_{-2} + S_{-3}
\nonumber\\
&=& S_{1,1} S_{-1} + S_{1} S_{-2} + S_{-3}
\\ [2mm]
S_{1,1,1} &=& \frac{1}{6} S_1^3 + \frac{1}{2} S_1 S_2 + \frac{1}{3}
S_{3} \nonumber\\ &=&
\frac{1}{3} \left[S_{-1,-1} S_{-1} + S_{-1} S_2 + S_{-3}\right]
\\ [2mm]
S_{2,1,1} + S_{1,2,1} + S_{1,1,2}
&=& \frac{1}{2} \left[S_1^2 + S_2 \right] S_{2} + S_1 S_{3} + S_{4}
\nonumber\\
&=& S_{1,1} S_{2} + S_1 S_{3} + S_{4}
\\ [2mm]
~~S_{2,-1,1} + S_{2,1,-1} + S_{1,2,-1} & &\nonumber\\
+ S_{1,-1,2} + S_{-1,2,1} + S_{-1,1,2} &=&
S_1 S_{-1} S_2 + S_2 S_{-2} + S_1 S_{-3}
+ S_{-1} S_{3} + 2 S_{-4}
\\ [2mm]
~~S_{-2,-1,1} + S_{-2,1,-1} + S_{1,-2,-1} & &\nonumber\\
+ S_{1,-1,-2} + S_{-1,-2,1} + S_{-1,1,-2} &=&
S_1 S_{-1} S_{-2} +  S_{-2}^2 + S_1 S_{3}
+ S_{-1} S_{-3} + 2 S_{4}
\\ [2mm]
S_{2,-1,-1} + S_{-1,2,-1} + S_{-1,-1,2}
&=& \frac{1}{2} \left[S_{-1}^2 + S_2 \right] S_{2} + S_{-1} S_{-3} 
+ S_{4}
\nonumber\\
&=& S_{-1,-1} S_{2} + S_{-1} S_{-3} + S_{4}
\\ [2mm]
S_{-2,1,1} + S_{1,-2,1} + S_{1,1,-2}
&=& \frac{1}{2} \left[S_1^2 + S_2 \right] S_{-2} + S_{-1} S_{-3} 
+ S_{-4}
\nonumber\\
&=& S_{1,1} S_{-2} + S_{-1} S_{-3} + S_{-4}
\\ [2mm]
S_{-2,-1,-1} + S_{-1,-2,-1} + S_{-1,-1,-2}
&=& \frac{1}{2} \left[S_{-1}^2 + S_{2} \right] S_{-2} + S_{-1} S_{3}
+ S_{-4}
\nonumber\\
&=& S_{-1,-1} S_{-2} + S_{-1} S_{3} + S_{-4}~.
\end{eqnarray}
Wherever possible we gave a second representation collecting part of
the terms into double sums which were given before.

Finally, the relations for sums with four indices read
\begin{eqnarray}
\sum_{{\rm perm}\{k,l,m,n\}} S_{k,l,m,n} &=& S_k S_l S_m S_n
+ \sum_{{\rm inv~perm}\{k,l,m,n\}} S_{k} S_{l}
S_{m \wedge n}
+ \sum_{{\rm inv~perm}\{k,l,m,n\}} 
S_{k \wedge l}
S_{m \wedge n} \nonumber\\ & &
+ 2 \sum_{{\rm inv~perm}\{k,l,m,n\}} 
S_{k} S_{l \wedge m \wedge n}
+ 6~S_{k \wedge l \wedge m \wedge n}~.
\end{eqnarray}
One obtains 5 other relations up to level four,
\begin{eqnarray}
S_{-1,-1,-1,-1} &=& 
\frac{1}{4} S_4 + \frac{1}{8} S_2^2  + \frac{1}{3}
S_{-3} S_{-1} + \frac{1}{4} S_2 S_{-1}^2 + \frac{1}{24} S_{-1}^4
\nonumber\\ &=&
\frac{1}{4} \left[S_{-1,-1,-1} S_{-1} + S_{-1,-1} S_2 + S_{-1} S_{-3}
+ S_4\right]
\\ [2mm]
~~S_{1,-1,-1,-1} + S_{-1,1,-1,-1} & & \nonumber\\
+ S_{-1,-1,1,-1} + S_{-1,-1,-1,1} &=& 
\frac{1}{6} S_1^3 S_{-1} + \frac{1}{3} S_{-1} S_3  
+ S_1 S_{-3} + S_{-4}
\nonumber\\ & & 
+ \frac{1}{2} \left[S_1^2 S_{-2} + S_1 S_{-1} S_2 + S_2 S_{-2}
\right] 
\nonumber\\ &=&
S_{1,1,1} S_{-1} + S_{1,1} S_{-2} + S_1 S_{-3} + S_{-4}
\\ [2mm]
~~S_{-1,-1,1,1} + S_{-1,1,-1,1} + S_{-1,1,1,-1}  & & \nonumber\\ 
+ S_{1,1,-1,-1} + S_{1,-1,1,-1} + S_{1,-1,-1,1}  &=&
S_{-1}^2 S_1^2 + S_{-1}^2 S_2 + 4 S_{-1} S_1 S_{-2} + S_2 S_1^2
\nonumber\\ & &
+ S_2^2 + 2 S_{-2}^2 + 4 S_{-1} S_{-3} + 4 S_1 S_3  +  S_4
\nonumber\\ &=&
4\left[ S_{1,1} S_{-1,-1} + S_{-2,-2} + S_{-1,-3} + S_{-3,-1} 
+ S_{1,3} \right.
\nonumber\\ & & \left.
+ S_{3,1} - S_4 \right]
\\ [2mm]
~~S_{-1,1,1,1} + S_{1,-1,1,1} & & \nonumber\\
+ S_{1,1,-1,1} + S_{1,1,1,-1} &=& \frac{1}{6} S_{-1}^3 S_1
+ \frac{1}{3} S_{-3} S_1 + S_{-1} S_3 + S_{-4} \nonumber\\ & &
+ \frac{1}{2} \left[ S_{-1} S_1 S_2 + S_{-1}^2 S_{-2}
+ S_{-2} S_2 \right]
\nonumber\\ &=& S_{-1,-1,-1} S_1 + S_{-1,-1} S_{-2} + S_{-1} S_3 +
S_{-4}
\\ [2mm]
S_{1,1,1,1} &=& 
\frac{1}{4} S_4 + \frac{1}{8} S_2^2  + \frac{1}{3}
S_3 S_1 + \frac{1}{4} S_2 S_1^2 + \frac{1}{24} S_1^4
\nonumber\\ &=&
\frac{1}{4} \left[S_{1,1,1} S_{1} + S_{1,1} S_2 + S_{1} S_{3}
+ S_4\right]~.
\end{eqnarray}
More relations are given in Ref.~\cite{JB1}.

The sums of the type $S_{-1,-1, \ldots,-1}$ can be evaluated
in compact form at arbitrary depth.
\footnotesize
\begin{eqnarray}
S_{\underbrace{\mbox{\scriptsize -1, \ldots ,-1}}_{\mbox{\scriptsize
$k$}}}
= \frac{1}{k!}
\left|\begin{array}{rrrrr}
 S_{-1}(N) &       1  &       0 & \ldots & 0      \\
-S_{2}(N) & S_{-1}(N) &       2 & \ldots & 0      \\
 S_{-3}(N) &-S_{2}(N) & S_{-1}(N)&\ldots & 0      \\
 \vdots   & \vdots   & \vdots  &         & \vdots \\
 (-1)^{k+1}S_{(-1)^k k}(N)  & (-1)^k S_{(-1)^{k-1} (k-1)}(N)&
 (-1)^{k-1} S_{(-1)^{k-2} (k-2)}(N) &
 \ldots & S_{-1}(N)
\end{array} \right| 
\end{eqnarray}
\normalsize
Similarly
the corresponding expressions for $S_{1,1,\ldots,1}$ are
\footnotesize
\begin{eqnarray}
S_{\underbrace{\mbox{\scriptsize 1, \ldots ,1}}_{\mbox{\scriptsize
$k$}}}
= \frac{1}{k!}
\left|\begin{array}{rrrrrr}
 S_{1}(N) &       1  &       0 & 0      & \ldots & 0      \\
-S_{2}(N) & S_{1}(N) &       2 & 0      & \ldots & 0      \\
 S_{3}(N) &-S_{2}(N) & S_{1}(N)& 3      & \ldots & 0      \\
 \vdots   & \vdots   & \vdots  & \vdots &        & \vdots \\
 (-1)^{k+1}S_{k}(N)  & (-1)^k S_{k-1}(N)& (-1)^{k-1} S_{k-2}(N) &
 (-1)^{k-2} S_{k-3}(N)& \ldots & S_1(N)
\end{array} \right|~.
\end{eqnarray}
\normalsize

which are given by
\begin{eqnarray}
S_{\underbrace{\mbox{\scriptsize -1, \ldots ,-1}}_{\mbox{\scriptsize
5}}}
&=&
 \frac{1}{120}\,     S_{-1}^{5}
+\frac{1}{12} S_{2}\,S_{-1}^{3}
+\frac{1}{6}\,S_{-3}\,S_{-1}^{2}
+\frac{1}{4}\,S_{4}\,S_{-1}
+\frac{1}{8}\,S_{-1}\,S_{2}^{2}
+\frac{1}{6}\,S_{2}\,S_{-3}
\nonumber\\ & &
+\frac{1}{5}\,S_{-5}
\label{eqsu1}
\\
S_{\underbrace{\mbox{\scriptsize -1, \ldots ,-1}}_{\mbox{\scriptsize
6}}}
&=&
 \frac{1}{720}\,      S_{-1}^{6}
+\frac{1}{48}\,S_{2}\,S_{-1}^{4}
+\frac{1}{18}\,S_{-3}\,S_{-1}^{3}
+\frac{1}{8}\, S_{4}  S_{-1}^2
+\frac{1}{5}\, S_{-5}\,S_{-1}
+\frac{1}{16}\,S_{-1}^{2} S_{2}^{2}
\nonumber\\ & &
+\frac{1}{6}\, S_{-1}\,S_{2}\,S_{-3}
+\frac{1}{48}\,S_{2}^{3}
+\frac{1}{8}\, S_{2}\,S_{4}
+\frac{1}{18}\,S_{-3}^{2}
+\frac{1}{6}\, S_{6}
\\
S_{\underbrace{\mbox{\scriptsize -1, \ldots ,-1}}_{\mbox{\scriptsize
7}}}
&=&
 \frac{1}{5040}\,S_{-1}^{7}
+\frac{1}{240}\, S_{-1}^{5} S_{2}
+\frac{1}{72}\,  S_{-1}^{4} S_{-3}
+\frac{1}{24}\,  S_{-1}^{3} S_{4}
+\frac{1}{10}\,  S_{-1}^{2} S_{-5}
+\frac{1}{6}\,   S_{-1}\,   S_{6}
\nonumber\\ & &
+\frac{1}{10}\,  S_{2}\,    S_{-5}
+\frac{1}{24}\,  S_{2}^{2} S_{-3}
+\frac{1}{48}\,  S_{2}^{2} S_{-1}^{3}
+\frac{1}{48}\,  S_{2}^{3} S_{-1}
+\frac{1}{12}\,  S_{-3}     S_{4}
+\frac{1}{18}\,  S_{-3}^{2} S_{-1}
\nonumber\\ & &
+\frac{1}{8}\,   S_{-1}\,   S_{2}\, S_{4}
+\frac{1}{12}\,  S_{-1}^{2} S_{2}\, S_{-3}
+\frac{1}{7}\,   S_{-7}
\\
S_{\underbrace{\mbox{\scriptsize -1, \ldots ,-1}}_{\mbox{\scriptsize
8}}}
&=&
 \frac{1}{40320}\,        S_{-1}^{8}
+\frac{1}{1440}\,         S_{-1}^{6} S_{ 2}
+\frac{1}{360}\,          S_{-1}^{5} S_{-3}
+\frac{1}{96}\,           S_{-1}^4   S_{ 4}
+\frac{1}{30}\,           S_{-1}^{3} S_{-5}
\nonumber\\ & &
+\frac{1}{12}\,           S_{-1}^{2} S_{ 6}
+\frac{1}{7}\,            S_{-1}     S_{-7}
+\frac{1}{36}\,           S_{-1}^{3} S_{2}\,   S_{-3}
+\frac{1}{16}\,           S_{-1}^{2} S_{2}\,   S_{4}
+\frac{1}{24}\,           S_{-1}\,   S_{2}^{2} S_{-3}
\nonumber\\ & &
+\frac{1}{12}\,           S_{-1}\,   S_{-3}\,   S_{4}
+\frac{1}{96}\,           S_{-1}^{2} S_{2}^{3}
+\frac{1}{36}\,           S_{-1}^{2} S_{-3}^{2}
+\frac{1}{384}\,          S_{2}^{4}
+\frac{1}{192}\,          S_{2}^{2}  S_{-1}^{4}
+\frac{1}{32}\,           S_{2}^{2}  S_{4}
\nonumber\\ & &
+\frac{1}{36}\,           S_{2}\,    S_{-3}^{2}
+\frac{1}{12}\,           S_{2}\,    S_{6}
+\frac{1}{10}\,           S_{2}\,    S_{-1}\,  S_{-5}
+\frac{1}{15}\,           S_{-3}\,   S_{-5} 
+\frac{1}{32}\,           S_{4}^{2}
+\frac{1}{8}\,            S_{8}~.
\label{eqsu2}
\end{eqnarray}
in explicit form beyond level~4.
The corresponding relations for $S_{\underbrace{\mbox{\scriptsize 1,
\ldots ,1}}_{\mbox{\scriptsize $n$}}}$ are obtained by substituting
the indices $-k \rightarrow k$ in Eqs.~(\ref{eqsu1}--\ref{eqsu2}).
Note the similarity to the corresponding sums of positive integer
powers studied by {\sc F. Fa\`{a} di Bruno}~\footnote{For historical
aspects see also~\cite{HIST}.}, \cite{FDB,RAMA}.

One may evaluate these sums also using the recursion relations
\begin{eqnarray}
S_{\underbrace{\mbox{\scriptsize -1, \ldots ,-1}}_{\mbox{\scriptsize
$k$}}} &=& \frac{1}{k} \sum_{l=1}^k S_{(-1)^l |l|} 
S_{\underbrace{\mbox{\scriptsize -1, \ldots ,-1}}_{\mbox{\scriptsize
$k-l$}}} \\
S_{\underbrace{\mbox{\scriptsize 1, \ldots ,1}}_{\mbox{\scriptsize
$k$}}} &=& \frac{1}{k} \sum_{l=1}^k S_{l} 
S_{\underbrace{\mbox{\scriptsize 1, \ldots ,1}}_{\mbox{\scriptsize
$k-l$}}} 
\end{eqnarray}

Sums of this type were studied already long ago~\cite{BINET,SCHEIBNER}.
A relation to derivatives of the $\Gamma$-function was given in
\cite{NIELSEN}. Recently a generating function for the sums
$S_{\underbrace{\mbox{\scriptsize 1, \ldots ,1}}_{\mbox{\scriptsize
$k$}}}$ has been derived~\cite{VERM1}
\begin{eqnarray}
(-1)^N x B(N+1,x-N) = 1 + \sum_{l=1}^{\infty} x^l
S_{\underbrace{\mbox{\scriptsize 1, \ldots ,1}}_{\mbox{\scriptsize
$l$}}}(N)~.
\end{eqnarray}
Finally we would like to mention a relation (cf.~\cite{ADAMC}) similar
to those for
$S_{\underbrace{\mbox{\scriptsize 1, \ldots ,1}}_{\mbox{\scriptsize
$k$}}}$ ralating the {\sc Stirling}--numbers of the first kind
to the finite harmonic sums of positive index,
\begin{eqnarray}
\left[\begin{array}{c} n \\ m \end{array}\right] = (n-1)
\left[\begin{array}{c} n-1 \\ m \end{array}\right]
+ \left[\begin{array}{c} n-1 \\ m-1 \end{array}\right] =
\frac{\Gamma(n)}
{\Gamma(m)} w(n,m-1),
\end{eqnarray}
with
\begin{eqnarray}
w(n,0) &=& 1 \nonumber\\
w(n,m) &=& \sum_{k=0}^{m-1} (1-m)_k S_{k+1}(n-1) w(n,m-1-k)
\end{eqnarray}
and $(r)_k = \prod_{l=1}^k (r+k-1)$ the {\sc Pochhammer--Barnes}
symbol.

The permutation relations lead to a first reduction of the number of 
direct {\sc Mellin}--transforms (\ref{eqME1}--\ref{eqME80}) and relate 
the transcendental functions occurring as arguments of the transforms
in the representations of section~\ref{sec:dirmellin}.

\subsection{Partial Permutations and Triple Harmonic Sums}
Aside of the permutation-invariant sum of harmonic sums over a given
index set one can consider equivalent decompositions of more symmetric
sums to obtain further algebraic relations. For triple harmonic sums
one may study the expression
\begin{eqnarray}
\label{eqT}
T_{a,b,c}(N) = \sum_{k=1}^N \frac{1}{k^a}
\left[\sum_{l=1}^k \frac{1}{l^b} \right]\left[
\sum_{m=1}^k \frac{1}{m^b}\right]~.
\end{eqnarray}
It was shown by {\sc Borwein} and {\sc Girgensohn}~\cite{BG1} for 
non-alternating finite sums, that $T_{a,b,c}(N)$~\footnote{In
ref.~\cite{BG1} relations between the sums $\eta_N(k,l,m) = S_{k,l,m}(N) 
- S_{k+l,m}(N) - S_{k,l+m}(N) + S_{k+l+m}(N)$ were discussed. We rewrite 
them in terms of the sums $S_{k_1,...,k_r}$ used throughout the present
paper.}  may have four representations in terms of triple harmonic sums. 
Earlier similar relations were derived by {\sc Markett}~\cite{MARKE} for 
infinite non-alternating sums. We generalize the relations allowing also 
for alternating  sums. These relations read
%
\begin{eqnarray}
\label{Ta}
T &=& S_{a,b,c} + S_{a,c,b} - S_{a \wedge b,c} - S_{a \wedge c,b}
     -S_{a,b \wedge c}+ S_{a \wedge b \wedge c} \\
T &=& S_c S_{a,b} - S_{c,a,b} + S_{c,a \wedge b} - S_c S_{a \wedge b} \\
T &=& S_b S_{a,c} - S_{b,a,c} + S_{b,a \wedge c} - S_b S_{a \wedge c} \\
\label{Td}
T &=& S_{b,c,a} + S_{c,b,a} - S_{b \wedge c, a}  - S_c S_{b,a}
     + S_b S_{a,c}  - S_b S_{a \wedge c}
\end{eqnarray}
They result into three relations between the harmonic sums for $a,b$ and 
$c$ being different and into two relations for two of the variables being
equal. The global permutation relation Eq.~(\ref{perm3}) is a consequence
of Eqs.~(\ref{Ta}--\ref{Td}).

Let us finally write down the relations for the cases discussed in
the present paper explicitly. The basis elements were choosen already
according to the results of the previous section, taking those sums
as basis elements which have the most simple functional structure
with respect to the argument of the {\sc Mellin}--transform. We have no
algebraic argument to predict this structure, which had to be found
by explicit analytic integration. To choose the most simple structures
as basis elements has the advantage that the analytic representations
of the {\sc Mellin}--transforms for complex values of $N$, 
cf.~\cite{JB3}, becomes as simple as possible.
For the level--3 sums the following relations hold
\begin{eqnarray}
S_{-1,-1,1} &=& \frac{1}{2}\left( -S_{-1,1,-1} +S_{-1}S_{-1,1}
+ S_{-1,-2} + S_{2,1}\right)\\
S_{1,-1,-1} &=& \frac{1}{2} \left(-S_{-1,1,-1} + S_{-1} S_{1,-1}
-S_{2,1}-S_{-1,-2}+S_1 S_2 + S_{-1} S_{-2} + 2 S_3 \right)\\
S_{-1,1,1}  &=& S_{1,1,-1} - S_1 S_{1,-1} + S_{-2,1} + S_{-1,2} + 
\frac{1}{2} \left(S_1^2 S_{-1}-S_{-1} S_2\right) -S_{-3}  \\
S_{1,-1,1}  &=& -2 S_{1,1,-1} + S_1 S_{1,-1} - S_{-2,1} - S_{-1,2}
+ S_1 S_{-2}  + S_{-1} S_2 + 2 S_{-3}~.
\end{eqnarray}
The relations for the threefold level--4 sums are~:
\begin{eqnarray}
S_{1,2,1} &=& - 2 S_{2,1,1} + S_{3,1} + S_1 S_{2,1} + S_{2,2} 
\\
S_{1,1,2} &=& S_{2,1,1} + \frac{1}{2} \left[ S_1 \left(S_{1,2} - S_{2,1}
              \right) + S_{1,3} - S_{3,1}\right] \\
S_{1,-2,1} &=& - 2 S_{-2,1,1} + S_{-3,1} + S_1 S_{-2,1} + S_{-2,2} \\
S_{1,1,-2} &=& S_{-2,1,1} + S_{-2} S_2 - S_{-2,2} - S_{-2} S_{1,1}
+ S_1 S_{1,-2} +S_{1,-3} - S_1 S_{-3}\\
S_{-1,-1,2} &=&  \frac{1}{2} \left( S_{-1} S_{-1,2} + S_{-1,-3} +
S_{2,2} - S_{-1,2,-1}\right)\\
S_{2,-1,-1} &=& \frac{1}{2} \left(-S_{-1,2,-1} + S_{-3,-1}
+ S_{-1} S_{2,-1} + S_{2,2}\right)\\
S_{-1,-1,-2} &=& \frac{1}{2}\left(-S_{-1,-2,-1}  + S_{2,-2}
+S_{-1} S_{-1,-2} + S_{-1,3}\right)\\
S_{-2,-1,-1} &=& \frac{1}{2}\left(-S_{-1,-2,-1} + S_{-2,2} + S_{3,-1}
+S_{-1} S_{-2,-1}\right)\\
S_{-2,-1,1} &=&  -S_{1,-2,-1} - S_{-2,1,-1} + S_{3,1} + S_{-3,-1}
+ S_4 \nonumber\\ & &
+ S_1 S_{-2,-1} + S_{1,3} - S_1 S_3 + S_{-2,-2}\\
S_{1,-1,-2} &=&  -S_{-1,-2,1} - S_{-1,1,-2} + S_{-1,-3}
+ S_{-2,-2} + S_1 S_{-1,-2}\\
S_{1,-2,-1} &=&  S_{-1,-2,1} - S_{-1} S_{-2,1} - S_{-1,-3} 
+ S_{-1} S_{-3} + S_1 S_{-2,-1} + S_{1,3} - S_1 S_3\\
S_{2,-1,1} &=&  -S_{1,2,-1} - S_{2,1,-1} + S_{3,-1} + S_2 \left(
S_{-1,1}+S_{1,-1}\right) \nonumber\\ & &
+ S_{2,-2}-S_2 S_{-2} -S_1 S_{-1,2} +S_1 S_{-3}\\
S_{1,-1,2} &=& -S_{1,2,-1}-S_{2,1,-1}+S_{1,-3}+S_{3,-1}+S_2 S_{1,-1}\\
S_{-1,1,2} &=&  S_{1,2,-1} + S_{2,1,-1} -S_{-1,2,1} + S_{-2,2} + S_{-3,1}
\nonumber\\ & &
-S_{-4} + S_{-1,3} - S_{3,-1} -S_2 S_{1,-1} + S_1 S_{-1,2} - S_1 S_{-3}
\end{eqnarray}

Using the algebraic relations given above one may show that the linear
representations up to threefold sums given in section~\ref{sec:dirmellin}
can be expressed by the single harmonic sums $S_{\pm k}(N)$ and the 
{\sc Mellin}--transforms of the functions
\begin{equation}
\begin{array}{cccc}
{\ds \frac{\log(1+x)}{x+1}} & {\ds \frac{\log^2(1+x)-\log^2(2)}{x-1}} &
{\ds \frac{\log^2(1+x)}{x+1}} &  \\
& & & \\
{\ds \frac{\Li_2(x)}{x+1}} & {\ds \frac{\Li_2(x)-\zeta(2)}{x-1}} &
{\ds \frac{\Li_2(-x)}{x+1}} & {\ds \frac{\Li_2(-x)-\zeta(2)/2}{x-1}} \\
& & & \\
{\ds \frac{\log(x)\Li_2(x)}{x+1}}&{\ds \frac{\log(x)\Li_2(x)}{x-1}} &
{\ds \frac{\Li_3(x)}{x+1}} & {\ds \frac{\Li_3(x)-\zeta(3)}{x-1}} \\
& & & \\
{\ds \frac{\Li_3(-x)}{x+1}} & {\ds \frac{\Li_3(-x)-3 \zeta(3)/4}{x-1}} &
{\ds \frac{\Sf(x)}{x+1}} & {\ds \frac{\Sf(x)-\zeta(3)}{x-1}} \\
& & & \\
{\ds \frac{\Sf(-x)}{x+1}} & {\ds \frac{\Sf(x^2)}{x+1}} &
{\ds \frac{\Sf(x^2)-\zeta(3)}{x-1}} & {\ds 
\log(1-x) \frac{\Li_2(-x)}{x+1}} \\
& & & \\
{\ds \frac{\log(1+x) - \log(2)}{x-1} \Li_2(x)} &
{\ds \frac{\log(1+x) - \log(2)}{x-1} \Li_2(-x)} &
{\ds \frac{\log(1-x) \Li_2(x)}{1+x}} &
{\ds \frac{\log(1+x)}{1+x} \Li_2(x)}\\
& & & \\
& & & \\
\end{array}
\label{funlist}
\end{equation}

\vspace{3mm}\noindent
According to {\sc Carlson's}--theorem~\cite{CARLS} the analytic 
continuation of {\sc Mellin}--transforms to complex numbers of $N$ can be
performed uniquely. The corresponding expressions for the 
{\sc Mellin}--transforms of the functions in (\ref{funlist}) are
given in Ref.~\cite{JB3}.

The {\sc Mellin}--transform of complicated functions as $F_1(x)$,
Eq.~(\ref{eqF1}), or {\sc Nielsen}--functions with more complicated
arguments as $(1\pm x)/2$, which occur in the triple sums with the
indices $\pm 1$, can be represented by polynomials of the
{\sc Mellin}--transforms of the functions listed above.
\section{Conclusions}
\label{sec:conclu}

\vspace{2mm}
\noindent
The {\sc Mellin}--transforms  of the {\sc Nielsen}--functions which emerge
in the splitting functions and {\sc Wilson}--coefficients in massless QED
and QCD up to two--loop order were calculated for all the contributing 
functions $f_i(x)$. They can be expressed by finite harmonic sums up to 
level four. The mathematical properties of {\sc Mellin}--transforms of 
this class can be understood easiest from the point of view of the 
alternating and non--alternating finite harmonic sums. These are 
polynomials over a substet of harmonic sums, which are 
related to {\sc Mellin}--transforms of {\sc Nielsen}--functions of 
relatively simple functional or argument structure. The latter properties
could only be revealed by the explicit calculation of all finite harmonic
sums up to level four in the linear representation. Comparing the linear
representation with the representations induced by the algebraic 
relations one may obtain a class of linear combinations of convolution 
relations between {\sc Nielsen}--functions. The fourfold level--four
sums are found to contribute to the splitting functions and 
{\sc Wilson}--coefficients up to two loop order only in their simplest 
variants, which can be represented algebraically in terms of lower level 
sums. They are, however, expected to contribute to the quantities beyond
the two--loop level. All remaining harmonic sums can be constructed out 
of about 20 basic {\sc Mellin}--transforms and the known representations 
which are expressible in terms of the function $\psi^{k}(z)$.
The splitting functions and coefficient functions depend usually only
on a subset of these harmonic sums requiring even less basic functions
for the representation of their {\sc Mellin}--transforms.

\vspace{2mm}
\noindent
{\bf Acknowledgement.}\\
We would like to thank W. van Neerven,
V. Ravindran and J. Vermaseren for useful discussions.
This 
work was supported in part by EC contract FMRX-CT-0194 (DG 12 - MIHT).

\newpage
\section{Appendix: Mellin Transforms}

\renewcommand{\arraystretch}{1.0}
\noindent
\begin{center}
\begin{tabular}{||r||c|l||}
\hline
\hline
& &   \\[-3mm]
\multicolumn{1}{||c||}{No.}&
\multicolumn{1}{c|}{${\ds f(z)}$}&
\multicolumn{1}{c||}{${\ds M[f](N) = \int_0^1 dz z^{N-1} f(z)}$}
\\
& &   \\[-3mm]
\hline
\hline
& &   \\[-3mm]
1 & ${\ds \delta(1-z)}$ & ${\ds 1}$   \\
& &   \\[-3mm]
\hline
& &   \\[-3mm]
2 & ${\ds z^{r}}$ & ${\ds \frac{1}{N+r}}$   \\
\hline
\hline
\end{tabular}
\end{center}

\noindent
\begin{center}
\begin{tabular}{||r||c|l||}
\hline
\hline
& &   \\[-3mm]
\multicolumn{1}{||c||}{No.}&
\multicolumn{1}{c|}{${\ds f(z)}$}&
\multicolumn{1}{c||}{${\ds M[f](N)}$}
\\
& &   \\[-3mm]
\hline
\hline
& &   \\[-3mm]
13 & ${\ds z^{r}\log(1+z)}$  &
${\ds \frac{(-1)^{N+r-1}}{N+r}\biggl[-S_{1}(N+r)+\frac{1+(-1)^{N+r-1}}
{2}}$   \\
 & & \\[-3mm]
& &${\ds\times S_{1}\biggl(\frac{N+r-1}{2}\biggr)+\frac{1-(-1)^{N+r-1}}{2}
S_{1}\biggl(\frac{N+r}{2}\biggr)\biggr]}$
\\
& &   \\
\hline
\hline
\end{tabular}
\end{center}
\pagebreak

\noindent
\begin{center}
\begin{tabular}{||r||c|l||}
\hline
\hline
& &   \\[-3mm]
\multicolumn{1}{||c||}{No.}&
\multicolumn{1}{c|}{${\ds f(z)}$}&
\multicolumn{1}{c||}{${\ds M[f](N)}$}
\\
& &   \\[-3mm]
\hline
\hline
& &  \\[-3mm]
20 & ${\ds \frac{1}{1-z}\log(z)\log(1-z)}$ &
${\ds \zeta(3)+\zeta(2)S_{1}(N-1)-S_{1}(N-1)S_{2}(N-1)-S_{3}(N-1)}$
\\
& & \\[-3mm]
\hline
& & \\[-3mm]
21 & ${\ds z^{r}\log(z)\log(1+z)}$ &
${\ds (-1)^{N+r}\frac{\zeta(2)}{2(N+r)}
  -\frac{1+(-1)^{N+r-1}}{(N+r)^{2}}\log(2)
  +\frac{(-1)^{N+r}}{N+r}}      $
\\
\hline
\hline
\end{tabular}
\end{center}
\pagebreak

\noindent
\begin{center}
\begin{tabular}{||r||c|l||}
\hline
\hline
& &   \\[-3mm]
\multicolumn{1}{||c||}{No.}&
\multicolumn{1}{c|}{${\ds f(z)}$}&
\multicolumn{1}{c||}{${\ds M[f](N)}$}
\\
& &   \\[-3mm]
\hline
\hline
&  & \\[-3mm]
24 & ${\ds \frac{1}{1-z}\log(z)\log^{2}\!(1-z)}$
& ${\ds S_{2}^{2}(N-1)-\zeta(2)S_{2}(N-1)+2S_{4}(N-1)}$  \\
& &  \\[-3mm]
& & ${\ds +S_{1}^{2}(N-1)
\bigl[S_{2}(N-1)-\zeta(2)\bigr]}$ \\
& &  \\
\hline
\hline
\end{tabular}
\end{center}
\pagebreak

\noindent
\begin{center}
\begin{tabular}{||r||c|l||}
\hline
\hline
& &   \\[-3mm]
\multicolumn{1}{||c||}{No.}&
\multicolumn{1}{c|}{${\ds f(z)}$}&
\multicolumn{1}{c||}{${\ds M[f](N)}$}
\\
& &   \\[-3mm]
\hline
\hline
& &  \\[-3mm]
30 & ${\ds \frac{1}{1+z}\log z\log^{2}\!(1+z)}$
& ${\ds 2 (-1)^{N} \Bigl[
 S_{1,1,-2}(N-1) + S_{1,2,-1}(N-1) + S_{2,1,-1}(N-1)}$  \\
& &  \\[-3mm]
& & ${\ds -S_{1,-2}(N-1)\log 2 - S_{2,-1}(N-1)\log 2
 + S_{1}(N-1)S_{2}(N-1)\log 2}$  \\
& &  \\
\hline
\hline
\end{tabular}
\end{center}
\pagebreak

\noindent
\begin{center}
\begin{tabular}{||r||c|l||}
\hline
\hline
& &   \\[-3mm]
\multicolumn{1}{||c||}{No.}&
\multicolumn{1}{c|}{${\ds f(z)}$}&
\multicolumn{1}{c||}{${\ds M[f](N)}$}
\\
& &   \\[-3mm]
\hline
\hline
& &  \\[-3mm]
32 & ${\ds z^{r}\log z\log(1+z)\log(1-z)}$
& ${\ds -\frac{(-1)^{N}}{N}\Bigl[
 S_{-2,1}(N) + S_{-1,2}(N) - \zeta(2)S_{-1}(N)}$  \\
& &  \\[-3mm]
& & ${\ds + \frac{13}{8}\zeta(3) - \frac{3}{2}\zeta(2)\log 2\Bigr]}$
\\
& &  \\
\hline
\hline
\end{tabular}
\end{center}
\pagebreak

\noindent
\begin{center}
\begin{tabular}{||r||c|l||}
\hline
\hline
& &   \\[-3mm]
\multicolumn{1}{||c||}{No.}&
\multicolumn{1}{c|}{${\ds f(z)}$}&
\multicolumn{1}{c||}{${\ds M[f](N)}$}
\\
& &   \\[-3mm]
\hline
\hline
& & \\[-3mm]
36 & ${\ds \frac{\Li_{2}(z)}{1+z}}$
 & ${\ds (-1)^{N-1}\biggl[ \zeta(2)S_{-1}(N-1) - S_{-2,1}(N-1)}$
 \\
& &  \\[-3mm]
& & ${\ds - \frac{5}{8}\zeta(3) + \zeta(2)\log 2\biggr]}$  \\
& &  \\
\end{tabular}
\end{center}
\pagebreak

\noindent
\begin{center}
\begin{tabular}{||r||c|l||}
\hline
\hline
& &   \\[-3mm]
\multicolumn{1}{||c||}{No.}&
\multicolumn{1}{c|}{${\ds f(z)}$}&
\multicolumn{1}{c||}{${\ds M[f](N)}$}
\\
& &   \\[-3mm]
\hline
\hline
& &  \\[-3mm]
43 & ${\ds z^{r}\Li_{2}(-z)\log z}$ &
${\ds \frac{(-1)^{N+r-1}}{(N+r)^{2}}
\biggl[\frac{2S_{1}(N+r)}{N+r}+S_{2}(N+r)\biggr]}$  \\
& &   \\[-3mm]
& & ${\ds-\frac{1+(-1)^{N+r-1}}{2(N+r)^{2}}\biggl[\frac{2}{N+r}
S_{1}\left(\frac{N+r-1}{2}\right)+\frac{1}{2}S_{2}\left(
\frac{N+r-1}{2}\right)}$  \\
& &   \\
\hline
\hline
\end{tabular}
\end{center}
\pagebreak

\noindent
\begin{center}
\begin{tabular}{||r||c|l||}
\hline
\hline
& &   \\[-3mm]
\multicolumn{1}{||c||}{No.}&
\multicolumn{1}{c|}{${\ds f(z)}$}&
\multicolumn{1}{c||}{${\ds M[f](N)}$}
\\
& &   \\[-3mm]
\hline
\hline
& & \\[-3mm]
47 & ${\ds z^{r}\Li_{2}(-z)\log(1+z)}$ &
${\ds \frac{(-1)^{N+r}}{N+r}\biggl\{\biggl[
 S_{2}(N+r)+\frac{\zeta(2)}{2}\biggr]\biggl[
\frac{1+(-1)^{N+r}}{2}S_{1}\left(\frac{N+r}{2}\right)}$ \\
& & ${\ds +\frac{1-(-1)^{N+r}}{2}S_{1}\left(\frac{N+r-1}{2}\right)
-S_{1}(N+r)\biggr]}$  \\
& & ${\ds-\biggl[\frac{1+(-1)^{N+r}}{4}S_{2}\left(\frac{N+r}{2}
\right)+\frac{1-(-1)^{N+r}}{4}S_{2}\left(\frac{N+r-1}{2}\right)}$
 \\
& & ${\ds -2S_{2}(N+r)\biggr]\log2}$  \\
\end{tabular}
\end{center}
\pagebreak

\noindent
\begin{center}
\begin{tabular}{||r||c|l||}
\hline
\hline
& &   \\[-3mm]
\multicolumn{1}{||c||}{No.}&
\multicolumn{1}{c|}{${\ds f(z)}$}&
\multicolumn{1}{c||}{${\ds M[f](N)}$}
\\
& &   \\[-3mm]
\hline
\hline
& & \\[-3mm]
49 & ${\ds \frac{\Li_{2}(-z)}{1+z}\log(1+z)}$ &
 ${\ds (-1)^{N-1}\biggl\{
 S_{1,2,-1}(N-1) + 2S_{2,1,-1}(N-1)}$  \\
& & ${\ds + \biggl[ S_{2,1}(N-1) - S_{1,-2}(N-1)
 - 2S_{2,-1}(N-1)}$  \\
& & ${\ds + S_{1}(N-1)S_{2}(N-1) + S_{3}(N-1)
 - \frac{1}{2}\zeta(2)S_{-1}(N-1)\biggr]\log 2}$  \\
& & ${\ds + \frac{1}{2}\zeta(2)S_{1,-1}(N-1)
 - \biggl[ S_{2}(N-1) - S_{-2}(N-1)\biggr]\log^{2}2}$  \\
\end{tabular}
\end{center}
\pagebreak

\noindent
\begin{center}
\begin{tabular}{||r||c|l||}
\hline
\hline
& &   \\[-3mm]
\multicolumn{1}{||c||}{No.}&
\multicolumn{1}{c|}{${\ds f(z)}$}&
\multicolumn{1}{c||}{${\ds M[f](N)}$}
\\
& &   \\[-3mm]
\hline
\hline
& &  \\[-3mm]
57 & ${\ds \tilde{\Phi}(z)}$ &
${\ds \frac{1}{N^{3}}+2\frac{(-1)^{N}}{N}\Bigl[S_{2}(N)
-\zeta(2)\Bigr]}$ \\
& & ${\ds -\frac{1+(-1)^{N}}{2N}\Bigl[S_{2}\left(\frac{N}{2}\right)
-\zeta(2)\Bigr]}$  \\
& & ${\ds +\frac{1-(-1)^{N}}{2N}\Bigl[S_{2}\left(\frac{N-1}{2}\right)
-\zeta(2)\Bigr]}$  \\
& &  \\
\hline
\hline
\end{tabular}
\end{center}
\pagebreak

\noindent
\begin{center}
\begin{tabular}{||r||c|l||}
\hline
\hline
& &   \\[-3mm]
\multicolumn{1}{||c||}{No.}&
\multicolumn{1}{c|}{${\ds f(z)}$}&
\multicolumn{1}{c||}{${\ds M[f](N)}$}
\\
& &   \\[-3mm]
\hline
\hline
& &  \\[-3mm]
64 & ${\ds \frac{\Li_{3}(-z)}{1+z}}$ &
  ${\ds (-1)^{N-1}\biggl\{
  S_{3,-1}(N-1) + \left[ S_{3}(N-1) - S_{-3}(N-1)\right]\log 2}$ \\
& & ${\ds + \frac{1}{2}\zeta(2)S_{-2}(N-1) -
  \frac{3}{4}\zeta(3)S_{-1}(N-1)}$  \\
& & ${\ds + \frac{1}{8}\zeta^{2}(2) -
  \frac{3}{4}\zeta(3)\log 2\biggr\}}$  \\
& &  \\
\hline
\hline
\end{tabular}
\end{center}
\pagebreak

\noindent
\begin{center}
\begin{tabular}{||r||c|l||}
\hline
\hline
& &   \\[-3mm]
\multicolumn{1}{||c||}{No.}&
\multicolumn{1}{c|}{${\ds f(z)}$}&
\multicolumn{1}{c||}{${\ds M[f](N)}$}
\\
& &   \\[-3mm]
\hline
\hline
& &  \\[-3mm]
70 & ${\ds z^{r}\Sf(z)}$ & ${\ds \frac{\zeta(3)}{N+r} -
\frac{S_{1}^{2}(N+r)+S_{2}(N+r)}{2(N+r)^{2}}}$ \\
& &  \\[-3mm]
\hline
& &  \\[-3mm]
71 & ${\ds \left(\frac{1}{1-z}\right)_{+}\Sf(z)}$ &
  ${\ds S_{2,1,1}(N-1) - \zeta(3)S_{1}(N-1) - \frac{6}{5}\zeta^{2}(2)}$
  \\
\hline\hline
\end{tabular}
\end{center}
\newpage

\end{document}